\documentstyle[12pt]{article}

\def\bra{\langle}
\def\ket{\rangle}

\begin{document}

\begin{center}
{\large \bf Recurrent dynamical symmetry breakings and
restorations\\ by Wilson lines at finite densities on a torus}
\end{center}
\medskip

\begin{center}
{Chung-Chieh Lee and Choon-Lin Ho \\ {\small \sl Department of
Physics, Tamkang University, Tamsui 25137, Taiwan}\\ ({\small Jan
27, 2000})}
\end{center}
\bigskip

\begin{abstract}
In this paper we derive the general expression of a one-loop
effective potential of the nonintegrable phases of Wilson lines
for an SU(N) gauge theory with a massless adjoint fermion defined
on the spactime manifold $R^{1,d-3}\times T^2$ at finite
temperature and fermion density. The Phase structure of the vacuum
is presented for the case with $d=4$ and $N=2$ at zero
temperature.  It is found that gauge symmetry is broken and
restored alternately as the fermion density increases, a feature
not found in the Higgs mechanism.  It is the manifestation of the
quantum effects of the nonintegrable phases.

\noindent PACS number(s): 11.10.Wx, 11.10.Kk, 11.15.Ex
\end{abstract}

\newpage
\section{Introduction}

By now no one would deny that one of our most important quests is
the search for the  theory that unifies all the known fundamental
interactions. This dream found its partial realization in the
Weinberg-Salam-Glashow model of electroweak theory.  From this
model one learns of the ironic yet important interplay between the
requirement in one's unified theory of certain gauge symmetries,
and the necessity of breaking them.

One of the most original and beautiful ideas in the search  of a
unified theory is the assumption of the existence of extra
compactified dimensions, originated from the works of Nordstr\"om,
Kaluza, and Klein \cite{ACF}. These compactified dimensions
manifest themselves as gauge fields in flat four-dimensional
spactimes.  The beauty and simplicity of this idea fascinated even
the great Einstein, who returned to it several times in the latter
part of his life \cite{EB}.  The idea, however, soon met with its
demise owing to the lack of experimental evidence.  It was not
until the 1970s that interest in Kaluza-Klein theories were
revived by modern attempts of unified theory such as the theories
of supergravity and superstrings.  In these modern theories,
however, gauge fields are usually assumed to be present in higher
dimensions. Nevertheless, it is fair to say that the principle of
gauge invariance, and the assumption of extra compactified
dimensions constitute the two most important ingredients in most
of the modern versions of a unified theory.  More recently, in a
different development, the possibility of extra dimensions has
also  attracted the attention of particle phenomenologists
\cite{ADD}:  it was argued that various unification scales might
be lower than what were previously thought of, if extra dimensions
exist.

Inspired by the Kaluza-Klein scenario, properties of gauge
theories in nontrivial spacetimes have also received considerable
interest during the last three decades.  Of the various topics
considered so far an interesting one is the study of the effects
of nonintegrable phases of the Wilson lines.   In a gauge theory
on a multiply connected space, these nonintegrable phases become
dynamical degrees of freedom \cite{Hoso,Toms}, and have many
interesting implications. They lead to a $\theta$-vacuum  in QED
on a circle \cite{circle}.  On toroidal spaces they induce
restriction on the Chern-Simons term \cite{CS}, and consequently
lead to multicomponent anyon wave functions and the related braid
group structures \cite{BG}, and quantum group symmetry \cite{Ho}.
Last but not least, they cause dynamical gauge symmetry breaking
in non-Abelian gauge theory \cite{Hoso,Toms}.

Dynamical gauge symmetry breaking by Wilson lines in a multiply
connected spacetime has been studied with considerable interests
since its introduction in the 1980s, and has been employed
extensively in superstring phenomenology \cite{GSW}. Unlike the
Higgs mechanism, which relies on the nonvanishing vacuum
expectation values of scalar fields, the Wilson line mechanism
depends essentially on the nontrivial holonomy of the vacuum gauge
field configuration arises from the topology of spacetime. An
interesting feature of this mechanism is that fermion mass
generation does not necessarily require symmetry to be broken, and
vice versa.

However, it is generally very difficult to compute dynamically the
quantum effects (in terms of the effective potentials) that
determine the symmetry-breaking patterns, except for a few simple
compact manifolds, such as the circles $R^{1,d-2}\times S^1$
\cite{Hoso,DM} and the torus $R^{1,d-3}\times T^2$
\cite{HeHo,McL1}.  So far no computation of the effective
potential has been achieved on the Calabi-Yau manifolds which are
relevant to superstring phenomenology. To gain a better
understanding of the mechanism, it is therefore useful to explore
various aspects of the mechanism in simple manifolds.

One interesting consideration is the inclusion of the effects of
finite temperature and finite fermion density.   Effect of finite
temperatures on Wilson line symmetry breaking has been studied for
circular \cite{S1,HoHoso} and toroidal \cite{McL2} compactified
spaces previously.  In \cite{LeeHo} both the effects of the
temperature and the fermion density are considered for the
circular case ($R^{1,d-2}\times S^1$).  These studies indicate
that in an SU(N) gauge theory with only gauge fields and fermions
in the adjoint representation of the group, higher temperatures
always restore the gauge symmetry (on the circle and torus,
symmetry is not broken by this mechanism when the fermions are in
the fundamental representation).

Our main purpose in this paper is to present the general
expressions of the one-loop effective potentials for the SU(N)
gauge fields and the adjoint fermions on the spacetime manifold
$R^{1,d-3}\times T^2$, with temperature and density effects taken
into account.   We also consider the $d=4$ dimensional case in
detail, and investigate the symmetry-breaking patterns as function
of the fermion density and the size of the torus at zero
temperatures. Even in this simple case an interesting feature
already emerges, namely, the gauge symmetry can be broken and
restored alternately as the density and/or the sizes of the torus
change. This is something not noted in the Higgs mechanism.

The organization of the paper is as follows.  First, we give a
brief review of the essence of the Wilson line mechanism in Sec.2.
Section 3 presents the derivation of the general expressions of
the effective potentials.  In Sec.4 the $d=4$ dimensional case is
studied numerically.  Section 5 concludes the paper.

\section{Wilson lines mechanism}

In this paper we consider an SU(N) gauge theory with adjoint
fermions on a $d$-dimensional spacetime manifold $R^{1,d-3}$
$\times T^2$.  Generalizing to the case in higher dimensional
torus is straightforward.  The Lagrangian is given by
\begin{equation} {\cal L}=-{1\over 2}~{\rm Tr} ~F_{\mu\nu}F^{\mu\nu}
+i\bar\psi \gamma^\mu D_\mu \psi~,
\end{equation}
where
\begin{eqnarray}
F_{\mu\nu}=\partial_\mu A_\nu-\partial_\nu
A_\mu+ig[A_\mu,A_\nu]~,\nonumber
D_\mu\psi=\partial_\mu\psi+ig[A_\mu,\psi]~.
\end{eqnarray}
As the space is multiplyconnected, one must specify the boundary
conditions of the fields for the two noncontractible loops. Let
$x^\alpha$ ($\alpha=0,1,\ldots,d-3$) and $y^a$ ($a=1,2$) label the
Minkowski and the toroidal coordinates, respectively, ($\mu,\nu$
run over both $\alpha$ and $a$).  The boundary conditions are
\begin{eqnarray}
&&A_\mu(x^\alpha,y^a+L_a)= U_a A_\mu(x^\alpha,y^a)U^\dagger_a~,\\
&&\psi (x^\alpha,y^a+L_a)= e^{i\beta_a}U_a\psi (x^\alpha,y^a)
U^\dagger_a~. \label{BC}
\end{eqnarray}
Here $L_a$ are the lengths of the circumferences of $T^2$.  The
phases $e^{i\beta_a}$ represent the continuous spin structure of
the manifold when using Dirac fermions.  These phases cancel from
physical operators constructed bilinearly in $\bar\psi$ and
$\psi$, but contribute to the boundary conditions.

One can then specify a particular SU(N) bundle over the torus, and
determine the vacuum configuration of the connection by evaluating
the effective potential for $\bra A_\mu \ket$. However, in this
paper, as in \cite{HeHo}, we will confine ourselves to the trivial
bundle with $\bra A_\mu \ket$ a constant element of SU(N). Even
this simple case gives interesting and nontrivial results on the
vacuum structure in the theory.

Suppose  that $\bra F_{\mu\nu} \ket=0$ in the vacuum and therefore
\begin{equation}
\bra A_\mu\ket =-{i\over g}V^{\dagger}\partial_\mu V  ~.
\end{equation}
This, in general, is physically distinct from $\bra A_\mu \ket =0$
in a multiplyconnected space, since there is no gauge
transformation which connects these two configurations without
spoiling the boundary conditions.   $V(x,y)$, which is
undetermined in classical theory,  is determined by quantum
effects as a function of the boundary conditions $(U_a,\beta_a)$
up to a global gauge transformation.  When the vacuum gauge
configuration $\bra A_\mu \ket$ is transformed to $\bra A'_\mu
\ket =0$, the boundary conditions $U_a$ are rotated to
$$U^{sym}_a=V(x^\alpha,y^a+L_a)U_a V^{\dagger}(x^\alpha,y^a)~.$$
As $U^{sym}_a$ are the boundary condition matrices in the gauge
$\bra A'_\mu \ket =0$, the residual gauge symmetry of the theory
is generated by those generators of the group which commute with
$U^{sym}_a$. This viewpoint of symmetry breaking was stressed by
Witten \cite{Witten}. One must, however, be reminded that the
above argument is valid only for $d\geq 3$.  In one-dimensional
space ($d=2$), $A_\mu$ does not take on definite values because of
quantum fluctuations, and therefore $V$ and $U^{sym}_a$ are not
well defined.

We note that $U^{sym}_a$ are related to the path-ordered Wilson
line integrals \begin{equation} W_a(x,y)=P
\exp{ig\int_{y^a}^{y^a+L_a} A\cdot dx}~U_a~.
\end{equation}
in that they have the same eigenvalues $e^{i\theta^a_i}$, the
nonintegrable phases.  These nonintegrable phases arise only in a
topologically nontrivial space and cannot be gauged away; their
values are determined dynamically.  The restriction to flat
connections greatly constrains the form of $\bra A_a \ket$ on the
torus (we assume $\bra A_\alpha \ket=0$ on the Minkowski space),
since all components must commute.  Hence in general $\bra A_a
\ket$ is a diagonal constant element of SU(N), and we define
\begin{eqnarray}
\bra A_a \ket_{ij} ={1\over gL} \theta^a_i \delta_{ij}~,~~
\sum^N_{i=1} \theta^a_i=0~~, \label{vac}
\end{eqnarray}
where $i,j$ are the SU(N) matrix indices.

One must evaluate the effective potential for $\bra A_\mu \ket$
(or equivalently for the nonintegrable phase $\theta^a_i$) in
order to determine the residual gauge symmetry of the system.  We
compute the one-loop effective potentials.  For this purpose, it
is convenient to take the background gauge with the gauge-fixing
function
\begin{eqnarray}
F[A]&=&  D_\mu[\bra A\ket]A^\mu\nonumber\\
    &=& \partial_\mu A^\mu + ig\left[\bra A_\mu\ket,A^\mu\right]\nonumber\\
    &=& 0~,
\end{eqnarray}
and the gauge parameter $\alpha=1$.  After performing an
analytical continuation to Euclidean spacetime, one obtains the
one-loop effective potential for the gauge and ghost fields
\begin{eqnarray}
V^{g+gh}_{eff}=+\frac{d-2}{2} \ln \det \left(-D^2\right)~,
\label{Vg}
\end{eqnarray}
and for the massless Dirac fermions
\begin{eqnarray}
V^{ad}_{eff}=-\frac{f(d)}{2} \ln \det \left(-D^2\right)~,
\label{Vf}
\end{eqnarray}
where $f(d)\equiv 2^{[d/2]}$ is the number of components of a
Dirac fermion. Here $[x]$ is the integral part of $x$.  In the
vacuum configuration (\ref{vac}), the operators $-D^2$ are given
by
\begin{equation}
\left[-D^2\right]_{jk}=-\partial_\alpha\partial^\alpha
-\sum_{a=1}^2\left[
\partial_a + \frac{i}{L_a}(\theta^a_j -\theta^a_k)\right]^2~.
\label{D2}
\end{equation}
We note here that the toroidal components of $-D^2$ act as mass
terms in the Minkowski space.   The effective potentials
(\ref{Vg}) and (\ref{Vf}) can be evaluated by the zeta function
method \cite{HeHo,HoHoso,LeeHo}, according to which one has
\begin{equation}
\ln \det (-D^2)=-\zeta^\prime (0)~,
\end{equation}
where the zeta function $\zeta (s)$ is defined according to the
field contents.  The next section is devoted to the evaluation of
the effective potentials.

\section{The one-loop effective potentials}

The one-loop effective potentials at finite temperature and
density can be evaluated according to the standard techniques in
finite-temperature field theory \cite{FT}.  The imaginary-time
formalism appropriate for the study of thermal equilibrium
properties is adopted here.  In this formalism the time coordinate
is Wick rotated to the Euclidean time $\tau=it$. The real time in
the time-evolution operator $\exp(-itH)$ for a Hamiltonian $H$ is
then related, by analytic continuation, to the inverse temperature
$\beta=1/T$ in the Boltzmann factor $\exp(-\beta H)$. The
(anti-)commutativity of bosonic (fermionic) fields then requires
the fields be (anti-)periodic in $\beta$.  The prescription of
deriving the effective potential of fermions at finite density is
to modify the Euclidean time derivative $\partial_\tau$ by
$\partial_\tau\to \partial_\tau -i\mu$, where $\mu$ is the
chemical potential of the fermions.  In the  following sections,
we present the results of applying the above techniques to the
evaluation of Eqs.(\ref{Vg}) and (\ref{Vf}).

\subsection{Gauge and ghost fields}

Evaluation of the one-loop effective potential of the gauge and
ghost fields is comparatively easy by following the steps in
\cite{HoHoso}.

The zeta function for the gauge and ghost fields is
\begin{eqnarray}
 \zeta^{g+gh}(s)
&&
=\frac{S_{d-3}}{\Gamma(s)}\int^\infty_0~dt~t^{s-1}\int_0^{\infty}\frac{dp\
 p^{d-4}}{(2\pi)^{d-3}}\frac{1}{\beta L_1L_2}
 \sum_{j,k}\sum_{n,m_1,m_2=-\infty}^{\infty}\nonumber\\
&&\times\exp\left[-t \left\{ p^2+ \left[ \frac{2\pi}{\beta}n
\right]^2
 +\left[ \frac{1}{L_{1}}(2\pi m_{1}+ \bar{\theta}^{1}_{jk})
\right]^2
 + \left[ \frac{1}{L_{2}}(2\pi m_{2}+ \bar{\theta}^{2}_{jk})
\right]^2 \right\}\right]~. \label{zeta-g}
\end{eqnarray}
Here $\bar{\theta}^a_{jk} \equiv \theta^a_{j}-
\theta^a_{k}~(a=1,2)$, $S_{d-3}\equiv
2\pi^{(d-3)/2}/\Gamma((d-3)/2)$ is the surface area of a
$(d-3)$-dimensional unit sphere, $\beta=1/T$ is the inverse
temperature, and $t$ is a dummy integration parameter not to be
confused with the time coordinate.

One can perform the $p$ and $t$ integrations in Eq.(\ref{zeta-g})
using the identity
\begin{eqnarray}
 \sum_{m_a=-\infty}^{\infty}
\exp\left\{-t\left[\frac{1}{L_a}(2\pi
m_{a}+\bar{\theta}^a_{jk})\right]^2\right\}
 = \frac{L_a t^{-\frac{1}{2}}}{\sqrt{4\pi}}\sum_{m_a=-\infty}^{\infty}
 \exp\left(-\frac{L_a^2m_a^2}{4t}+i m_a\bar{\theta}^a_{jk}\right)~,
\label{iden1}
\end{eqnarray}
which can be proven by means of the Poisson sum formula. Keeping
only the finite part as $s\to 0$, one arrives at the effective
potential for the gauge and ghost fields:
\begin{eqnarray}
 V^{g+gh}_{eff}(T)
 &=& -\frac{(d-2)\Gamma(d/2)}{\pi^{d/2}}
 \sum_{j,k} \left\{
 \sum_{m_1=1}^{\infty}\frac{\cos (m_1\bar{\theta}^1_{jk})}{L_1^dm_1^d}
 + \sum_{m_2=1}^{\infty}\frac{\cos
(m_2\bar{\theta}^2_{jk})}{L_2^dm_2^d}\right.\nonumber\\
 &+& 2\sum_{m_1=1}^{\infty}\sum_{m_2=1}^{\infty}
  \frac{\cos (m_1\bar{\theta}^1_{jk})\cos (m_2\bar{\theta}^2_{jk})}
 {(L_1^2m_1^2+L_2^2m_2^2)^{d/2}}\nonumber\\
 &+& 2\sum_{n=1}^{\infty}
\left[  \sum_{m_1=1}^{\infty}
 \frac{\cos (m_1\bar{\theta}^1_{jk})}{(\beta^2n^2+L_1^2m_1^2)^{d/2}}
 + \sum_{m_2=1}^{\infty}
 \frac{\cos (m_2\bar{\theta}^2_{jk})}{(\beta^2n^2+L_2^2m_2^2)^{d/2}}\right.
\nonumber \\ &+& \left.\left.
2\sum_{m_1=1}^{\infty}\sum_{m_2=1}^{\infty}
  \frac{\cos (m_1\bar{\theta}^1_{jk})\cos (m_2\bar{\theta}^2_{jk})}
 {(\beta^2n^2+L_1^2m_1^2+L_2^2m_2^2)^{d/2}}\right]\right\}~.
\end{eqnarray}

\subsection{Adjoint fermionic field}

The zeta function for the adjoint fermion fields is
\begin{eqnarray}
\zeta^{ad}(s)
=&&\frac{S_{d-3}}{\Gamma(s)}\int^\infty_0~dt~t^{s-1}\int_0^{\infty}\frac{dp\
 p^{d-4}}{(2\pi)^{d-3}}\frac{1}{\beta L_1L_2}\nonumber\\
&&\times \sum_{j,k}\sum_{n,m_1,m_2=-\infty}^{\infty} \exp\left[-t
\left\{ p^2+ \left[ \frac{2\pi}{\beta}(n+\frac{1}{2})+i\mu
\right]^2 \right.\right. \nonumber\\
 &+&\left.\left.  \left[ \frac{1}{L_{1}}(2\pi m_{1}+
\hat{\theta}^{1}_{jk}) \right]^2
 + \left[ \frac{1}{L_{2}}(2\pi m_{2}+ \hat{\theta}^{2}_{jk})
\right]^2 \right\}\right]~, \label{zeta-f}
\end{eqnarray}
where $\mu$ is the chemical potential of the fermions, and
$\hat{\theta}^a_{jk} \equiv \theta^a_{j}-
\theta^a_{k}-\beta_a~(a=1,2)$. Evaluation of this zeta function is
done by generalizing the steps in \cite{S2,LeeHo}.

Performing the $p$ integration with the help of Eq.(\ref{iden1})
and the identity
\begin{eqnarray}
 \sum_{n=-\infty}^{\infty}
 \exp\left\{-t\left[ \frac{2\pi}{\beta}(n+\frac{1}{2})+i\mu \right]^2\right\}
 = \frac{\beta}{\sqrt{4\pi}}t^{-\frac{1}{2}}
 \left[ 1 +2\sum_{n=1}^{\infty} (-1)^{n}\cosh (n\beta \mu)
 e^{-\frac{\beta^2n^2}{4t}} \right]~,
\end{eqnarray}
one obtains the finite part of the effective potential as follows:
\begin{eqnarray}
V^{ad}_{eff}
 &=& \frac{f(d)}{2^{d-1}\pi^{\frac{d-2}{2}}L_1L_2}
 \int_{0}^{\infty}dt~ t^{-\frac{d}{2}}\nonumber\\
&&\times\sum_{j,k}  \left\{ \sum_{m_1,m_2=-\infty}^{\infty}
\exp\left[-t\left\{ \left[\frac{1}{L_1}(2\pi
m_1+\hat{\theta}^{(1)}_{jk})\right]^2
   +\left[\frac{1}{L_2}(2\pi m_2+\hat{\theta}^{(2)}_{jk})\right]^2
\right\}\right]\right. \nonumber\\ &+& 2\sum_{n=1}^{\infty}
(-1)^{n}\cosh (n\beta \mu)
  \sum_{m_1,m_2=-\infty}^{\infty}
 \exp\left[-t\left\{ \left[\frac{1}{L_1}(2\pi
m_1+\hat{\theta}^{(1)}_{jk})\right]^2\right.\right. \nonumber\\
&&\left.\left.\left. \ \ \ \   +\left[\frac{1}{L_2}(2\pi
m_2+\hat{\theta}^{(2)}_{jk})\right]^2 \right\}
-\frac{\beta^2n^2}{4t}\right] \right\}~. \label{vf1}
\end{eqnarray}
The first term on the right-hand side (rhs) of Eq.(\ref{vf1}) is
independent of the temperature $T$.

Making use of the identity
\begin{eqnarray}
 \int_{0}^{\infty}dt ~t^{-\frac{d}{2}-1}
 e^{-tM^2-\frac{\beta^2n^2}{4t}}
 = 2 \left( \frac{2M}{n\beta}\right)^{\frac{d}{2}}K_{d/2}(n\beta M)~,
\end{eqnarray}
we now integrate Eq. (\ref{vf1}) with respect to $t$ to get
\begin{eqnarray}
V_{eff}^{ad}(T,\mu)=&&\nonumber\\
 \frac{f(d)\Gamma (d/2)}{\pi^{d/2}}
 \sum_{j,k}&& \left\{
 \sum_{m_1=1}^{\infty}\frac{\cos (m_1\hat{\theta}^1)}{L_1^dm_1^d}
 + \sum_{m_2=1}^{\infty}\frac{\cos (m_2\hat{\theta}^2)}{L_2^dm_2^d}
 + 2\sum_{m_1=1}^{\infty}\sum_{m_2=1}^{\infty}
  \frac{\cos (m_1\hat{\theta}^1)\cos (m_2\hat{\theta}^2)}
 {(L_1^2m_1^2+L_2^2m_2^2)^{d/2}}
 \right\}\nonumber \\
&+& \frac{f(d)}{2^{d-3}\pi^{(d-2)/2}L_1L_2}
 \sum_{j,k}  \sum_{n=1}^{\infty} (-1)^{n}\cosh (n\beta \mu)\nonumber\\
&&\times\sum_{m_1,m_2=-\infty}^{\infty}
 \left(\frac{2M_{jk}^{m_1m_2}}{n\beta}\right)^{(d-2)/2}
 K_{(d-2)/2}\left(n\beta M_{jk}^{m_1m_2}\right)~,
\label{vf2}
\end{eqnarray}
where
\begin{eqnarray}
 M_{jk}^{m_1m_2}= \left\{\left[\frac{2\pi
m_1+\hat{\theta}^1}{L_1}\right]^2
        +\left[\frac{2\pi m_2+\hat{\theta}^2}{L_2}\right]^2
        \right\}^{\frac{1}{2}}~.
\end{eqnarray}
This is the general expression of the effective potential for the
adjoint fermion fields at finite temperature and density.

To facilitate the numerical analysis of the symmetry-breaking
patterns, it is realized that a different representation of the
femionic effective potential is desirable.  To this end we shall
transform Eq. (\ref{vf2}) into an integral form by using the
following integral representation of the modified Bessel function:
\begin{eqnarray}
 K_{\nu}(z) =
 \frac{\sqrt{\pi}}{\Gamma(\nu+\frac{1}{2})}\left(\frac{z}{2}\right)^{\nu}
 \int_1^\infty e^{-zx}(x^2-1)^{\nu-\frac{1}{2}}dx
 \ \ ; \ \
 Re(z) >0~,~Re(\nu) > -\frac{1}{2}
\end{eqnarray}
and summing over $n$.  This leads to
\begin{eqnarray}
V_{eff}^{ad}(T,\mu)&=&
 \frac{f(d)\Gamma (d/2)}{\pi^{d/2}}
 \sum_{j,k}\left[
 \sum_{m_1=1}^{\infty}\frac{\cos (m_1\hat{\theta}^1)}{L_1^dm_1^d}
 + \sum_{m_2=1}^{\infty}\frac{\cos
(m_2\hat{\theta}^2)}{L_2^dm_2^d}\right.\nonumber\\
 &+&\left. 2\sum_{m_1=1}^{\infty}\sum_{m_2=1}^{\infty}
  \frac{\cos (m_1\hat{\theta}^1)\cos (m_2\hat{\theta}^2)}
 {(L_1^2m_1^2+L_2^2m_2^2)^{d/2}}
 \right]
-\frac{f(d)}{2^{d-2}\pi^{(d-3)/2}\Gamma(\frac{d-1}{2})}\frac{1}{L_1L_2}
\nonumber \\ &&\times\sum_{j,k}  \sum_{m_1,m_2=-\infty}^{\infty}
 \left(M_{jk}^{m_1m_2}\right)^{d-2}
 \left[ \int_{1}^{\infty}
 \frac{(x^2-1)^{\frac{d-3}{2}}}{e^{\beta (x M_{jk}^{m_1m_2}-\mu)}+1}dx
 + \left(\mu\to -\mu\right) \right]~.
\label{vf3}
\end{eqnarray}

Equation (\ref{vf3}) is valid for those values of $m_1,m_2$ such
that $M_{jk}^{m_1m_2}\neq 0$. In the case where $M_{jk}^{m_1m_2}=
0$, a different representation is in order.  In this case, the
terms with these particular sets of $\{m_1,m_2\}$ in the second
term of the rhs. of Wq. (\ref{vf1}) become
\begin{eqnarray}
 &&\frac{f(d)}{2^{d-2}\pi^{\frac{d-2}{2}}}\frac{1}{L_1L_2}
 \sum_{n=1}^{\infty} (-1)^{n}\cosh (n\beta \mu)
 \int_{0}^{\infty}dt ~t^{-\frac{d}{2}}e^{-\frac{\beta^2n^2}{4t}}
 \bigskip \nonumber\\
 &=&\frac{f(d)}{2} \frac{\Gamma(\frac{d-2}{2})}{\pi^{\frac{d-2}{2}}}
 \frac{1}{L_1L_2\beta^{d-2}}
 \left[ \sum_{n=1}^{\infty}\frac{(-e^{\beta \mu})^{n}}{n^{d-2}}
 + \sum_{n=1}^{\infty}\frac{(-e^{-\beta \mu})^{n}}{n^{d-2}}
 \right]
 \bigskip \nonumber\\
 &=&\frac{f(d)}{2} \frac{\Gamma(\frac{d-2}{2})}{\pi^{\frac{d-2}{2}}}
 \frac{1}{L_1L_2\beta^{d-2}}
 \left[ \ \mbox{Li}_{d-2}(-e^{\beta \mu})
 + \mbox{Li}_{d-2}(-e^{-\beta \mu})\
 \right]~,
\label{Li}
\end{eqnarray}
where $\mbox{Li}_s (x)=\sum_{n=1}^\infty x^n/n^s$ is the
polylogarithmic function of order $s$ \cite{Lewin}.  This result
can also be obtained by using the asymptotic form of the modified
Bessel function
\begin{equation}
K_\nu(x)\approx 2^{\nu-1}\Gamma(\nu)\frac{1}{x^\nu}~, ~~~~x\to 0^+
\end{equation}
in Eqs.(\ref{vf2}) for those terms with $M_{jk}^{m_1m_2}= 0$
\cite{LeeHo}. To ensure better convergence in numerical
computation, we express the term $\mbox{Li}_{d-2}(-e^{\beta \mu})$
in terms of $\mbox{Li}_{d-2}(-e^{-\beta \mu})$ by means of the
identities\cite{Lewin}
\begin{eqnarray}
 \mbox{Li}_{s}(-x) + (-1)^{s}\mbox{Li}_{s}(-1/x)
 = -\frac{1}{s!}\ln^{s}(x)
 +2
 \sum_{r=1}^{[s/2]}\frac{\ln^{s-2r}(x)}{(s-2r)!}\mbox{Li}_{2r}(-1)
\end{eqnarray}
and
\begin{eqnarray}
 \mbox{Li}_{2r}(-1) = -\frac{(2^{2r-1}-1)}{(2r)!}\pi^{2r}B_r~.
\end{eqnarray}
Here  $B_r$ are the Bernoulli numbers.

\section{$SU(2)$ Gauge Theory on $R^{1,1}\times T^2$ Manifold
$(T=0,\mu \neq 0)$}

With the expressions of the effective potentials of the gauge
fields, the ghost fields, and the adjoint fermions given in the
last section, we can study the symmetry structures of the vacuum
at different temperatures and densities by looking at the values
of the nonintegrable phases $\theta_j^a$ which minimize the total
effective potential $V_{eff}=V_{eff}^{g+gh}+V_{eff}^{ad}$. As the
expressions are rather complicated, the computational task is
quite involved for large dimensionality $d$ and gauge group SU(N).
Interesting features, however, already surface even for small $d$
and $N$.  In this section, we study the symmetry patterns of an
SU(2) gauge theory with adjoint fermions in $d=4$ dimensions
numerically with finite densities at zero temperatures. The system
with zero density and finite temperature were considered
previously in \cite{McL2}.  Symmetry is always restored at high
temperatures.  We see that as the fermion density increases, the
gauge symmetry of the system  is broken and restored alternately.

In the case of SU(2) theory, there are only two independent
nonintegrable phases owing to the traceless conditions:
\begin{eqnarray}
 \theta^a_1 = - \theta^a_2 = \theta^a
 \ \ , \ \
 a = 1,2~.
\end{eqnarray}
For convenience, we set $r\equiv L_2/L_1$.

The effective potential for the gauge and ghost fields with fixed
$(\theta^a, r)$ now reads
\begin{eqnarray}
 V^{g+gh}_{eff}(\theta^a,r)
 = &-&\frac{4}{\pi^2L_{1}^4} \left\{
 \sum_{m_1=1}^{\infty}\frac{\cos 2m_1\theta^1}{m_1^4}
 + \sum_{m_2=1}^{\infty}\frac{\cos
2m_2\theta^2}{r^4m_2^4}\right.\nonumber\\
  &+& \left. 2\sum_{m_1=1}^{\infty}\sum_{m_2=1}^{\infty}
 \frac{\cos(2m_{1}\theta^1)\cos(2m_{2}\theta^2)}
 {(m_1^2+r^2m_2^2)^2}
 \right\}~.
\end{eqnarray}

For the fermionic fields, we may use
\begin{equation}
\frac{1}{e^{\beta x}+1}\to \theta(-x)~,
\end{equation}
where $\theta(x)$ is the Heaviside step function, to get
\begin{eqnarray}
&&V^{ad}_{eff} =\frac{4}{\pi^{2}L_1^4} \left\{ \left[
\sum_{m_1=1}^{\infty}\left(
 \frac{\cos m_1(2\theta^1-\beta_1)}{m_1^4}
+ \frac{\cos m_1(2\theta^1+\beta_1)}{m_1^4}
\right)\right.\right.\nonumber\\ && \ \ \ \ \ \ +
\sum_{m_2=1}^{\infty}\left(
 \frac{\cos m_2(2\theta^2-\beta_2)}{r^4m_2^4}
 + \frac{\cos m_2(2\theta^2+\beta_2)}{r^4m_2^4}
 \right)
\nonumber \\ &&\ \ \ \ \ \ +
2\sum_{m_1=1}^{\infty}\sum_{m_2=1}^{\infty}\left(
 \frac{\cos m_1(2\theta^1-\beta_1)
 \cos m_2(2\theta^2-\beta_2)}
 {(m_1^2+r^2m_2^2)^2}\right.\nonumber\\
&&\left. \left.\ \ \ \ \ \ + \frac{\cos m_1(2\theta^1+\beta_1)
 \cos m_2(2\theta^2+\beta_2)}
 {(m_1^2+r^2m_2^2)^2} \right) \right]
 \nonumber \\
 &&-\frac{\pi}{2r}\left[
 \sum_{m_1=0}^{\infty}\sum_{m_2=0}^{\infty}
\left( \ \left[\left(2\pi m_1+2\theta^1-\beta_1\right)^2
 + \left(\frac{2\pi m_2+2\theta^2-\beta_2}{r}\right)^2 \right]
 \int_{1}^{\frac{\mu L_1}{\sqrt{\left[ \cdots \right]}}}
 (x^2-1)^{\frac{1}{2}}dx\right.\right.
 \nonumber\\
&&\left. +\left[\left(2\pi m_1-2\theta^1-\beta_1\right)^2
 + \left(\frac{2\pi m_2-2\theta^2-\beta_2}{r}\right)^2 \right]
 \int_{1}^{\frac{\mu L_1}{\sqrt{\left[ \cdots \right]}}}
 (x^2-1)^{\frac{1}{2}}dx
 \ \right) \nonumber\\
&&+ \sum_{m_1=1}^{\infty}\sum_{m_2=0}^{\infty}
 \left( \ \left[\left(2\pi m_1-2\theta^1+\beta_1\right)^2
 + \left(\frac{2\pi m_2+2\theta^2-\beta_2}{r}\right)^2 \right]
 \int_{1}^{\frac{\mu L_1}{\sqrt{\left[ \cdots \right]}}}
 (x^2-1)^{\frac{1}{2}}dx\right.
\nonumber\\ &&\left. +\left[\left(2\pi
m_1+2\theta^1+\beta_1\right)^2
 + \left(\frac{2\pi m_2-2\theta^2-\beta_2}{r}\right)^2 \right]
 \int_{1}^{\frac{\mu L_1}{\sqrt{\left[ \cdots \right]}}}
 (x^2-1)^{\frac{1}{2}}dx
 \ \right) \nonumber\\
&&+ \sum_{m_1=0}^{\infty}\sum_{m_2=1}^{\infty} \left( \
\left[\left(2\pi m_1+2\theta^1-\beta_1\right)^2
 + \left(\frac{2\pi m_2-2\theta^2+\beta_2}{r}\right)^2 \right]
 \int_{1}^{\frac{\mu L_1}{\sqrt{\left[ \cdots \right]}}}
 (x^2-1)^{\frac{1}{2}}dx\right.
  \nonumber\\
&&\left. +\left[\left(2\pi m_1-2\theta^1-\beta_1\right)^2
 + \left(\frac{2\pi m_2+2\theta^2+\beta_2}{r}\right)^2 \right]
 \int_{1}^{\frac{\mu L_1}{\sqrt{\left[ \cdots \right]}}}
 (x^2-1)^{\frac{1}{2}}dx
 \ \right) \nonumber\\
&&+ \sum_{m_1=1}^{\infty}\sum_{m_2=1}^{\infty}
 \left( \ \left[\left(2\pi m_1-2\theta^1+\beta_1\right)^2
 + \left(\frac{2\pi m_2-2\theta^2+\beta_2}{r}\right)^2 \right]
 \int_{1}^{\frac{\mu L_1}{\sqrt{\left[ \cdots \right]}}}
 (x^2-1)^{\frac{1}{2}}dx\right.
\nonumber\\
 &&\left.\left.\left. +\left[\left(2\pi m_1+2\theta^1+\beta_1\right)^2
 + \left(\frac{2\pi m_2+2\theta^2+\beta_2}{r}\right)^2 \right]
 \int_{1}^{\frac{\mu L_1}{\sqrt{\left[ \cdots \right]}}}
 (x^2-1)^{\frac{1}{2}}dx
 \ \right)\right]~\right\}.
\label{vf4}
\end{eqnarray}
The symbol $[\cdots]$  in the upper limit of each integral in Eq.
(\ref{vf4}) represents the factor inside the square bracket
immediately in front of the respective integral.  In case this
factor equals zero, and the corresponding term in Eq.(\ref{vf4})
is replaced by
\begin{eqnarray}
- \frac{\mu^2}{\pi L_1^2 r}
\end{eqnarray}
obtained from Eq. (\ref{Li}).

For simplicity we only consider the boundary conditions
$\beta_1=\beta_2=0$. The vacuum of the system at fixed $\mu$ and
$r$ is determined by finding the values of the nonintegrable
phases $(\theta^1,\theta^2)$ which correspond to the global
minimum of the total one-loop effective potential $V_{eff}$. These
values of the nonintegrable phases in turn, through the Wilson
line integrals, determine the residual gauge symmetry of the
system, as described in Sect. 2.   Therefore, the vacuum and its
symmetry are determined dynamically. In the present case, the
global minima are found to be located at one of the following sets
of the possible values of $\theta^a$:
\begin{eqnarray}
(\theta^1,\theta^2)=(0,0),~(0,\pi/2),~(\pi/2,0)~~{\rm or}
~(\pi/2,\pi/2)~ (mod~\pi)~.
\end{eqnarray}
Of these locations only the case $(0,0)~(mod~\pi)$ represents
unbroken SU(2) symmetry.  We present the results as a phase
diagram with $\mu L_1$ versus $r=L_2/L_1$ in Fig.1.  One sees that
the gauge symmetry is broken and restored alternately as the
fermion density $\mu$ increases at fixed $r$, or as the size of
the torus changes at fixed density.  This is a new feature not
noticed so far in the Higgs mechanism, and is the manifestation of
the quantum effect of the nonintegrable phases $\theta^a$. In the
usual Higgs models,  gauge symmetry is usually restored at zero
temperature as the fermion density increases \cite{Higgs}.  Only
in gauge theories with neutral currents could an increase in
fermion density increase the symmetry breaking \cite{Linde}.
However, no Higgs model, as far as we know, exhibits such a
recurrent pattern of symmetry breaking and restoration.
Implications of this feature of the Wilson line mechanism in
particle and string phenomenology, and in astroparticle physics
have yet to be explored.

In the limit $r=L_2/L_1\to \infty$, the spacetime $R^{1,1}\times
T^2$ becomes the manifold $R^{1,2}\times S^1$. From Fig.~1 we see
that the recurrent pattern of symmetry breakings and restorations
persists in this limit.  Wilson lines symmetry breaking on the
spacetime $R^{1,2}\times S^1$ was studied in \cite{LeeHo} at
finite temperatures and densities.  The recurrent pattern of
symmetry breakings and restorations were, however, over looked  in
\cite{LeeHo} as we did not extend the range of $\mu$ far enough.
We therefore take this opportunity to present in Fig.~2 the
correct phase diagram for the SU(2) theory with the boundary
condition $\beta_1=0$ at finite temperatures and densities.  For
future reference we record here the first few critical values of
$\mu$ at $T=0$ that define the boundaries of symmetric and broken
phases: $\mu_c L_1(T=0)= 1.979, 6.146, 9.140, 12.495,
15.55,\ldots$.

\section{Conclusions}

In this paper we discuss Wilson line symmetry breaking of SU(N)
gauge theory with adjoint fermions on the $d$-dimensional
spacetime $R^{1,d-3}\times T^2$. General expressions of the
one-loop effective potentials of the gauge and ghost fields, and
the adjoint fermion fields were presented. Symmetry patterns of
the vacuum structure in a $d=4$ dimensional SU(2) theory at zero
temperature is considered in detail.   It  is noted in this case
that the gauge symmetry can be broken and restored alternately as
the fermion density changes.  We expect this result to be true for
all SU(N) groups and dimensionality $d$.  This is a new feature
not observed in the Higgs mechanism, and is the manifestation of
the quantum effects of the nonintegrable phases.

\vskip 1truein \centerline{\bf ACKNOWLEDGMENT}

This work was supported in part by the Republic of China through
Grant No. NSC 89-2112-M-032-004.

\newpage
\centerline{\bf Figure Captions}
\begin{description}
\item[Figure 1.] Symmetry patterns of the vacuum in an SU(2)
gauge theory on spacetime $R^{1,1}\times T^2$ with boundary
conditions $\beta_1=\beta_2=0$ at zero temperature and finite
density $\mu$.  The phase diagram is plotted as a function of $\mu
L_1$ and $r=L_2/L_1$, where $L_1$ and $L_2$ are the lengths of the
torus. Shaded regions represent the symmetric phase, while the
unshaded ones represent the broken phase.
\item[Figure 2.]  Symmetry patterns of the vacuum in an SU(2)
gauge theory on spaectime $R^{1,2}\times S^1$ with boundary
condition $\beta_1=0$ at finite temperature $T$ and density $\mu$.
The phase diagram is plotted as a function of $\mu L$ and $TL$,
where $L$ is the length of the circle.
\end{description}
\end{document}